\documentclass[a4paper,10pt]{article}

\usepackage{amssymb}
\usepackage{amsmath}
\usepackage{hyperref}

\title{Technical Report: Modelling Multiple Cell Types with Partial Differential Equations }
\author{Simon Tanaka \\ CoBi group (Prof. Iber), D-BSSE, ETH Zurich \\ \href{mailto:simon.tanaka@bsse.ethz.ch}{simon.tanaka@bsse.ethz.ch} }

\begin{document}

\maketitle

\begin{abstract}
Partial differential equations are a convenient way to describe reaction-advection-diffusion processes of signalling models.
If only one cell type is present, and tissue dynamics can be neglected, the equations can be solved directly.
However, in case of multiple cell types it is not always straight forward to integrate a continuous description of the tissue dynamics.
Here, we discuss (delayed) differentiation of cells into different cell types and hypertrophic cell volume change upon differentiation.
\end{abstract}

\section{Formulation}

Partial differential equations have been widely used to describe spatial signalling models in morphogenesis \cite{Turing1952a}.
Oftentimes only one cell type is assumed, and therefore the governing equations apply to the entire domain.
However, if the equations shall be dependent on multiple cell types, the latter have to be represented explicitly.
An obvious approach would be to model cell types as \textit{cell densities} or \textit{cell concentrations}, i.e. a scalar field for each cell type.
This approach has been chosen for example to model early bone development \cite{Tanaka2013}.
However, there are a few drawbacks associated with that approach.
Having cell mixtures may not be what you want.
Sharply defined domains with homogeneous cell populations are often more desirable.
It is also unnecessary to have two scalar fields to describe two cell types if they are complementary.
Let's assume that we have a cell type $\mathcal{C}_{1}$ differentiating into cell type $\mathcal{C}_{2}$.
The negation $\mathcal{C}_{1}^{\neg})$ should by definition be $\mathcal{C}_{2}$, and they are supposed to sum up to $\mathcal{C}_{1}+\mathcal{C}_{2}=1$ at any time and everywhere.

Alternatively, we can define the cell types $\mathcal{C}_{1}$ and $\mathcal{C}_{2}$ by means of an index scalar field $\mathcal{D}=\mathcal{D}\left(\boldsymbol{x},t\right)$:
\begin{equation}\label{eq:celldefinition}
\begin{array}{lcl}
\mathcal{C}_{1} & := & 1 \cdot (\mathcal{D}>\theta_{\mathcal{D}}) \\
\mathcal{C}_{2} & := & 1 \cdot (\mathcal{D} \leq \theta_{\mathcal{D}})
\end{array}
\end{equation}

where $\theta_{\mathcal{D}}$ is an arbitrary threshold value which may be set to $\theta_{\mathcal{D}}=1/2$.
To be consistent (see later), we also define the negations:
\begin{equation}
\begin{array}{lcl}
\mathcal{C}_{1}^{\neg} & := & 1 \cdot (\mathcal{D} \leq \theta_{\mathcal{D}}) \\
\mathcal{C}_{2}^{\neg} & := & 1 \cdot (\mathcal{D} > \theta_{\mathcal{D}})
\end{array}
\end{equation}

Obviously, the cell types can only take values $\in\{0,1\}$ such that the scalar variables $\mathcal{C}$ and $\mathcal{C}^{\neg}$ can be used in a simple way in signalling model equations.
For example, assuming that only $\mathcal{C}_{1}$ expresses the protein $\mbox{P}$, and only $\mathcal{C}_{2}$ can degrade it, we would write:
\begin{equation}
\partial_{t} \text{P}  + \nabla \cdot \left(\text{P} \mathbf{u} \right) = D_{\text{P}} \Delta \text{P} + \rho_{\text{P}} \cdot \mathcal{C}_{1} - \delta_{\text{P}} \cdot \text{P} \cdot \mathcal{C}_{2}
\end{equation}
where $D_{\text{P}}$ denotes the diffusion coefficient, $\mathbf{u}$ the velocity field in case of a growing or moving tissue, $\rho_{\text{P}}$ the production rate and $\delta_{\text{P}}$ the degradation rate.

The index field $\mathcal{D}$ is governed by the following equation:
\begin{equation}\label{eq:dummyequation}
\partial_{t} \mathcal{D} + \boldsymbol{u} \cdot \nabla \mathcal{D} =
D_{\mathcal{D}} \Delta \mathcal{D} -
\delta_{\mathcal{D}} \cdot \mathcal{D} \cdot \Xi \cdot \mathcal{C}_{1} -
\delta_{\mathcal{D}} \cdot \mathcal{D} \cdot \mathcal{C}_{1}^{\neg}
\end{equation}

Let's discuss the meaning of all terms.
The equation is of reaction-diffusion type with advection, but without dilution.
Note that there is no such thing as \textit{cell diffusion} in this model at the time; the diffusion is only for numerical purposes and shall be chosen as small as possible.
Further development of the model would be needed to include chemotaxis, cell migration and so on.
The condition function $\Xi$ is controlling the degradation of $\mathcal{D}$,
and it depends on the specific model.
It is potentially a boolean operator, i.e. once certain conditions are fulfilled, differentiation takes place.
For example, let's assume a hypothetical morphogen $\text{M}$ and assume that differentiation occurs whenever its concentration drops below a certain threshold $\theta_{\text{M}}$.
The condition function would read:
\begin{equation}
\Xi\left(\text{M}\right) =
\begin{cases}
1, & \mbox{if } \left(\text{M} < \theta_{\text{M}} \right) \\
0, & \mbox{otherwise }
\end{cases}
\end{equation}

The decay rate $\delta_{\mathcal{D}}$ controls how fast $\mathcal{C}_{1}$ is converted into $\mathcal{C}_{2}$, which will be discussed in Section \ref{sec:delay} in more detail.
The rightmost term in Eq. (\ref{eq:dummyequation}) is there for numerical reasons: the decay of $\mathcal{D}$ ceases right upon reaching the threshold $\theta_{\text{M}}$ as defined in Eq. (\ref{eq:celldefinition}).
That's why it is desirable to diverge from the critical threshold $\theta_{\text{M}}$, which is achieved by an arbitrary decay $\delta_{\mathcal{D}}$ controlled by $\mathcal{C}_{1}^{\neg}$.

The model can be extended by adding differentiation delay, growth (proliferation and volume increase upon differentiation) and differentiation cascades with multiple cell types which will be discussed in the following sections.

\section{Differentiation Delay}\label{sec:delay}

Since we have exponential decay of $\mathcal{D}$ in Equation (\ref{eq:dummyequation}),
the time to reach the critical value (in this case $\theta_{\mathcal{D}}=1/2$) can be easily computed:
\begin{equation}\label{eq:delaytime}
\tau^{\text{diff}} = \frac{\ln\left(2\right)}{ \delta_{\mathcal{D}}}
\end{equation}

This is the time span from reaching the differentiation triggering condition $\Xi$ to the actual cell type conversion.
The higher $\delta_{\mathcal{D}}$, the more instantaneous the differentiation.
To be more precise, it is a time integrator: The differentiation triggering condition has to be fulfilled for an accumulated time $\tau^{\text{diff}}$ before the actual cell type conversion takes place.
If an approximately instantaneous differentiation is desired, a large value of $\delta_{\mathcal{D}}$ is chosen such that $\tau^{\text{diff}}$ is negligibly small.
However, if the model requires a temporal delay, Eq. (\ref{eq:delaytime}) can be used to do so.


\section{Cell Volume Increase upon Differentiation}

One way of describing the mechanical properties of embryonic tissue is by assuming that it behaves similar to a viscous fluid on longer time scales.
This approach has been used to model limb bud \cite{Dillon1999} and embryonic long bone development \cite{Tanaka2013}.
The incompressible Navier-Stokes equations read:

\begin{subequations}\label{eq:navierstokes}
\begin{align}
\rho \left( \partial_{t} \boldsymbol{u} + \left( \nabla \cdot \boldsymbol{u} \right) \boldsymbol{u} \right) &=
		-\nabla p + \mu \left( \Delta \boldsymbol{u} + \frac{1}{3}\nabla\left(\nabla\cdot \boldsymbol{u}\right) \right) + \boldsymbol{f} \label{eq:momentumequation}\\
\rho \nabla \cdot \boldsymbol{u} &= \mathcal{S}
\end{align}
\end{subequations}
where $\rho$ is a constant mass density, $\boldsymbol{u}$ the velocity field, $p$ the pressure field, $\boldsymbol{f}$ an external force field, and $\mathcal{S}$ a local mass source to model cell proliferation and hypertrophic increase of cell volume.
This tissue model represents the mechanical properties of \textit{all} cell types.

Here, we seek to combine the viscous tissue model and our cell type model.
In a two dimensional setting, let's assume that the cells $\mathcal{C}_{1}$ are growing $\Phi$-fold in area when differentiating into $\mathcal{C}_{2}$.
This process might represent hypertrophic cell volume increase upon differentiation \cite{Tanaka2013}.
The naive attempt would be to simply write for the mass source:
\begin{equation}
\mathcal{S}^{\text{diff}} = (\Phi-1) \frac{\delta_{\mathcal{D}}}{ln\left(2\right)} \cdot
								\Xi \cdot \mathcal{C}_{1}
\end{equation}

That is, we have a net mass gain of $(\Phi-1)$ units (gaining $+\Phi$ and losing $-1$).
The rate $\delta_{\mathcal{D}} / ln\left(2\right)$ is the inverse from Eq. (\ref{eq:delaytime}), i.e., this mass gain has to take place within the differentiation time $\tau^{\text{diff}}$.
Of course, the same conditions hold as for the index field decay: the differentiation condition $\Xi$ has to be fulfilled \textit{and} we must have $\mathcal{C}_{1}$.

However, this leads to spurious results because the mass source leads to an expansion of the differentiation zone and thus to a systematic exceeding of the added mass.
We do not want to change the dynamics of $\mathcal{D}$ because we still want to control the differentiation time $\tau^{\text{diff}}$ with $\delta_{\mathcal{D}}$.
So the way to go is to correct the mass source such that the total mass increase is linear and not exponential.
However, we also do not want to integrate the domain areas or introduce an additional diluted factor (and make $\mathcal{S}^{\mathrm{diff}}$ proportional it).
But we can estimate the (accumulated) time $\tilde{\tau}$ a certain location is already differentiating by taking the local value of $\mathcal{D}$:
\begin{equation}
\tilde{\tau}\left(\boldsymbol{x},t\right) = -\frac{1}{\delta_{\mathcal{D}}} ln\left(\mathcal{D}\left(\boldsymbol{x},t\right)\right)
\end{equation}

Now let's consider the evolution of a small area $A$ with a scaled uniform mass source $\tilde{\mathcal{S}} = \mathcal{S}_{\mathrm{diff}} \cdot A_{0}/A(t)$:
\begin{equation}
\partial_{t} A = A \mathcal{S}_{\mathrm{diff}} \frac{A_{0}}{A} = \mathcal{S}_{\mathrm{diff}} A_{0}
\end{equation}

which can be integrated over the differentiation time $\tilde{\tau}$:
\begin{equation}
A = A_{0} + \mathcal{S}_{\mathrm{diff}} A_{0} \tilde{\tau}
\end{equation}

such that we find for our true scaled mass source $\tilde{\mathcal{S}}$:
\begin{equation}\label{eq:scaledsource}
\tilde{\mathcal{S}}\left(\tilde{\tau}\right) = \frac{\mathcal{S}_{\mathrm{diff}}} {1+\mathcal{S}_{\mathrm{diff}} \tilde{\tau}}
\end{equation}


A drawback of this formulation of the mass source $\tilde{\mathcal{S}}$ might be that it is spatially non-uniform in the differentiating zone.
Furthermore, the dependency of the source on the differentiation time $\tilde{\tau}$ represents an additional complexity and source of numerical errors.
Therefore, we derive a constant, effective source $\tilde{\mathcal{S}}^{c}$. 
Integration of the scaled source (cf. Eq. (\ref{eq:scaledsource})) leads to:
\begin{equation}
\int_{\tilde{\tau}=0}^{\tilde{\tau}=\frac{ln(2)}{\delta_{\mathcal{D}}}}
\frac{\mathcal{S}^{\text{diff}}} {1+\mathcal{S}^{\text{diff}} \tau} d\tilde{\tau}
=
ln\left( 1 + \mathcal{S}^{\text{diff}} \frac{ln(2)}{\delta_{\mathcal{D}}} \right)
\end{equation}

and the average, constant, uniform mass source $\tilde{\mathcal{S}}^{c}$ is determined as:
\begin{equation}
\tilde{\mathcal{S}}^{c} = \frac{\delta_{\mathcal{D}}}{ln(2)} ln\left( 1 + \mathcal{S}^{\text{diff}} \frac{ln(2)}{\delta_{\mathcal{D}}} \right)
\end{equation}


\section{Multiple Cell Types}\label{sec:multiplecelltypes}

Suppose we would like to have the following differentiation cascade: $\mathcal{C}_{1}$ $\rightarrow$ $\mathcal{C}_{2}$ $\rightarrow$ $\mathcal{C}_{3}$ $\rightarrow$ $\mathcal{C}_{4}$.
We need three index fields $\mathcal{D}_{1,2,3}$, and the cells are defined as:
\begin{equation}\label{eq:multicelldefinition}
\begin{array}{lcl}
\mathcal{C}_{1} & := & 1 \cdot (\mathcal{D}_{1}>\theta_{\mathcal{D}}) \\
\mathcal{C}_{2} & := & 1 \cdot (\mathcal{D}_{1} \leq \theta_{\mathcal{D}}) \cdot (\mathcal{D}_{2} > \theta_{\mathcal{D}}) \\
\mathcal{C}_{3} & := & 1 \cdot (\mathcal{D}_{1} \leq \theta_{\mathcal{D}}) \cdot (\mathcal{D}_{2} \leq \theta_{\mathcal{D}}) \cdot (\mathcal{D}_{3} > \theta_{\mathcal{D}}) \\
\mathcal{C}_{4} & := & 1 \cdot (\mathcal{D}_{1} \leq \theta_{\mathcal{D}}) \cdot (\mathcal{D}_{2} \leq \theta_{\mathcal{D}}) \cdot (\mathcal{D}_{3} \leq \theta_{\mathcal{D}})
\end{array}
\end{equation}

and the negations read:
\begin{equation}
\begin{array}{lcl}
\mathcal{C}_{1}^{\neg} & := & 1 \cdot (\mathcal{D}_{1} \leq \theta_{\mathcal{D}}) \\
\mathcal{C}_{2}^{\neg} & := & 1 \cdot (\mathcal{D}_{2} \leq \theta_{\mathcal{D}}) \\
\mathcal{C}_{3}^{\neg} & := & 1 \cdot (\mathcal{D}_{3} \leq \theta_{\mathcal{D}}) \\
\mathcal{C}_{3}^{\neg} & := & 1 \cdot (\mathcal{D}_{3} > \theta_{\mathcal{D}})
\end{array}
\end{equation}

Of course, by exploiting the idea of the binary system, the 4 cell types could also be represented by only two index fields, but that would make the logical operators a little bit less intuitive for now.

The index fields obey equations similar to Eq. (\ref{eq:dummyequation}).
The differentiation is one-way, i.e. the reversion is not possible, and e.g. differentiation into $\mathcal{C}_{4}$ is only possible iff we have $\mathcal{C}_{3}$ \textit{and} the differentiation condition is fulfilled.
If de-differentiation is really needed, the model would have to be extended.\\


\bibliographystyle{unsrt}
\bibliography{MyCollection.bib}

\end{document}